\documentclass[pdflatex,bst/sn-aps,twocolumn]{sn-jnl}

\usepackage{graphicx}%
\usepackage{multirow}%
\usepackage{amsmath,amssymb,amsfonts}%
\usepackage{amsthm}%
\usepackage{mathrsfs}%
\usepackage[title]{appendix}%
\usepackage{xcolor}%
\usepackage{textcomp}%
\usepackage{manyfoot}%
\usepackage{booktabs}%
\usepackage{algorithm}%
\usepackage{algorithmicx}%
\usepackage{algpseudocode}%
\usepackage{listings}%
\usepackage{geometry}
\usepackage{dsfont}
\usepackage{braket}
\usepackage{comment}

\usepackage{booktabs}%
\usepackage{float}%
\usepackage{bm}
\usepackage{todonotes}

\usepackage{graphicx}
\usepackage{url}

\theoremstyle{thmstyleone}%
\theoremstyle{thmstyletwo}%

\theoremstyle{thmstylethree}%

\begin{document}

\title{The surprising simplicity of forecasting quantum chaotic many-body dynamics}



\author[1,2]{\fnm{F.} \sur{Perciavalle}}
\equalcont{These authors contributed equally to this work.}

\author[3]{A. Gallo}
\equalcont{These authors contributed equally to this work.}

\author[1,2]{F. Plastina}

\author[3]{G. Greco}

\author[1,2]{N. Lo Gullo}

\author[3]{C. Adornetto}

\affil[1]{Dipartimento di Fisica, Università della Calabria, Rende 87036, Italy}

\affil[2]{INFN, Sezione LNF, gruppo collegato di Cosenza}

\affil[3]{Department of Mathematics and Computer Science, University of Calabria, Rende 87036, Italy}


\abstract{Accurately simulating chaotic quantum many-body dynamics remains a major computational challenge for classical methods, due to the rapid buildup and spatial spreading of entanglement during the evolution. This raises the question of whether machine learning can provide an effective alternative for predicting quantum dynamics. We address this question by formulating quantum dynamics as a time-series forecasting problem, using a few-qubit PXP chain, realizable with Rydberg-atom arrays, as a benchmark. By varying the initial state, the system spans dynamical regimes ranging from ergodic behavior to quantum many-body scarring, providing a controlled setting for testing forecasting models across qualitatively different dynamics. We compare two contrasting architectures: an expressive nonlinear Transformer and DLinear, a simple linear forecasting model. The Transformer accurately predicts dynamics in the more ergodic regime, but its performance progressively deteriorates as the initial state approaches the scarred limit. In contrast, DLinear remains accurate across the entire family of initial states, with its main deviations consisting of small high-frequency oscillations that have little effect on the overall prediction error. Remarkably, these results show that observables generated by complex quantum many-body dynamics can be forecast with high accuracy through a simple linear mapping from past to future observations. This reveals that the complexity of the underlying quantum evolution need not translate into an equally complex forecasting problem.}




\maketitle

\section{Introduction}
Simulating the dynamics of quantum many-body systems remains a fundamental computational challenge because the dimension of the Hilbert space grows exponentially with the size of the system~\cite{feynman1982simulating, georgescu2014quantum, daley2022practical, altman2021quantum}. Exact numerical methods therefore become rapidly intractable, motivating approximate approaches that exploit the structure of physically relevant quantum states. Among these, matrix product states (MPS) provide an efficient representation of states with limited entanglement and form the basis for powerful tensor-network methods for many-body simulations ~\cite{cirac2021matrix, verstraete2008matrix, paeckel2019time, benenti2018principles}. However, as nonequilibrium dynamics generates entanglement that progressively spreads across the system, an accurate MPS representation requires increasingly large computational resources. This limitation is particularly severe in quantum chaotic many-body systems, where entanglement typically grows rapidly after a generic quench and eventually approaches volume-law scaling~\cite{dalessio2016from, kim2013ballistic, bertini2019entanglement}.Exact diagonalization is thus limited by system size, whereas tensor-network methods are ultimately limited by the accessible evolution time.

\begin{figure*}[!t]
    \centering
    \includegraphics[width=0.95\linewidth]{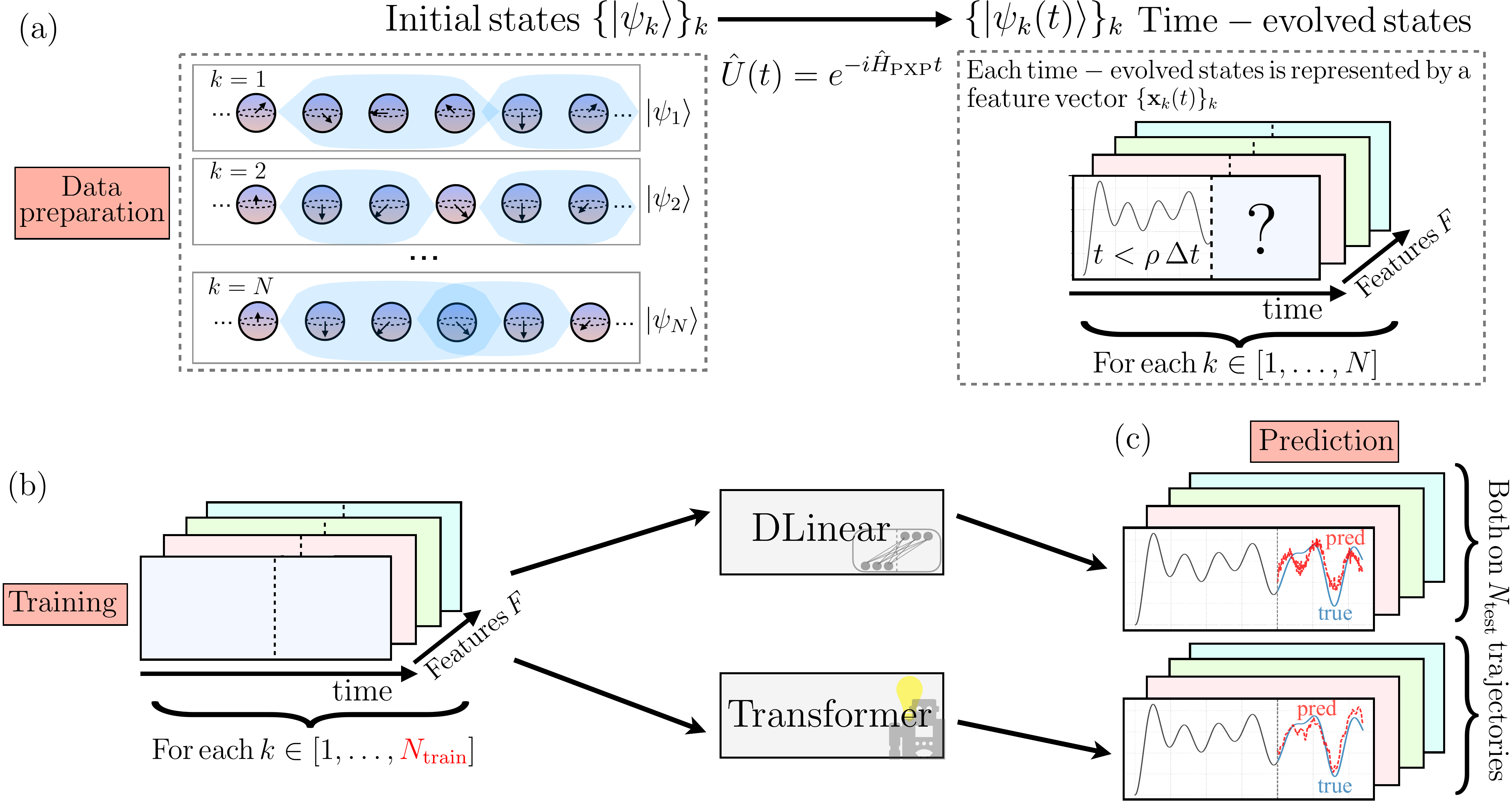}
    \caption{ Schematic illustration of the forecasting protocol, consisting of three stages: \textit{data preparation} [panel (a)], \textit{training} [panel (b)], and \textit{prediction} [panel (c)]. Data preparation (a): A set of $N$ initial quantum states, $\{\ket{\psi_k}\}_{k=1}^{N}$, of an $L$-qubit system is prepared. Each state is evolved under the Hamiltonian $\hat{H}_{\rm PXP}$, generating $N$ quantum trajectories, $\{\ket{\psi_k(t)}\}_{k=1}^{N}$. At each time $t$, each trajectory is represented by a feature vector $\mathbf{x}_k(t)$ comprising local magnetizations and connected two-point correlations, for a total of $F=L(L+1)/2$ features. For each trajectory, the first $\rho$ time steps define the input sequence, while the subsequent $\tau$ time steps constitute the target sequence. Training (b): DLinear and the Transformer are trained on a subset of $N_{\rm train}<N$ trajectories. Prediction (c): The trained models are evaluated on the remaining $N_{\rm test}=N-N_{\rm train}$ unseen trajectories to assess their forecasting performance.}

    \label{fig:sketch_}
\end{figure*}

The computational limitations of conventional simulation methods have stimulated a growing interest in data-driven approaches to quantum dynamics. In particular, machine-learning (ML) methods provide a complementary route to extracting and predicting dynamical behavior directly from generated or experimentally accessible data. Here, we consider precisely this forecasting setting, where the problem can be formulated in terms of observable time series: given a finite history of observations, the goal is to predict their subsequent evolution without explicitly propagating the full many-body quantum state. In this context, a variety of ML architectures have been explored, including long short-term memory (LSTM) networks, multilayer perceptrons (MLPs) and convolutional neural networks (CNNs) ~\cite{mazza2021machine, mohseni2024deep,pugzlys2025autoregressive, coppola2025learning}, as well as data-driven methods based on dynamical mode decomposition~\cite{kaneko2025forecasting}. 

A complementary perspective arises when the roles of quantum dynamics and machine learning are reversed: rather than using ML to predict quantum time evolution, quantum systems can themselves be exploited as computational resources for learning and predicting classical nonlinear and chaotic dynamics. As an example, in quantum reservoir computing the intrinsic dynamical complexity of quantum systems is harnessed for information processing and temporal prediction~\cite{fujii2021quantum,salatino2026forecasting,settino2025memory,sannia2026temporal,ramone2026quantum,sannia2024dissipation}. Thus, quantum dynamics can be viewed both as a target of ML and as a resource for learning complex dynamical patterns.

This dual perspective raises a natural question: when quantum dynamics are themselves the object to be learned, how much model complexity is actually required to predict them? More specifically, to what extent does the complexity of the underlying quantum evolution translate into the complexity required to forecast its observable dynamics?

In this work, we address this question by comparing two forecasting architectures with markedly different structures: a nonlinear Transformer model~\cite{vaswani2017attention, wang2026deep, kim2025comprehensive}, which uses self-attention to capture dependencies in sequential data, and a DLinear architecture~\cite{vaswani2017attention, wang2026deep, kim2025comprehensive}, a simple linear model originally developed for long-horizon time-series forecasting. This comparison allows us to directly investigate whether forecasting complex quantum dynamics requires an expressive nonlinear architecture, or whether a simple linear mapping from past to future observations is sufficient to accurately predict the evolution of a quantum many-body system.

The quantum many-body system considered in this work is the paradigmatic PXP model~\cite{bernien2017probing,bluvstein2021controlling, serbyn2021quantum, turner2018quantum, liang2025observation,dutta2025prethermalization, hudomal2022driving}, an effective model describing interacting Rydberg atoms in the blockade regime. Its dynamics depend strongly on the choice of initial state, allowing qualitatively different dynamical regimes to be explored within the same Hamiltonian. In particular, the model exhibits both ergodic dynamics and non-ergodic evolution associated with quantum many-body scars (QMBS), characterized by long-lived coherent revivals~\cite{turner2018quantum, serbyn2021quantum, chandran2023quantum, ren2025scar}. By considering a family of initial states interpolating between these regimes, we can therefore assess the forecasting performance of linear and nonlinear models across different types of quantum dynamics while keeping the underlying Hamiltonian fixed.

Our results reveal a clear difference between the two forecasting approaches for this problem. DLinear provides accurate and stable predictions across the entire family of initial states considered, whereas the performance of the Transformer depends strongly on the dynamical regime, progressively deteriorating as the dynamics approach the quantum-scarred limit. The robust performance of DLinear is particularly notable given its simple linear structure, showing that, for the observable dynamics considered here, accurate forecasting can be achieved without an expressive nonlinear architecture. This finding is consistent with recent observations in classical time-series forecasting~\cite{zeng2023transformers}, where simple linear models have been shown to compete with, and in some cases outperform, more sophisticated nonlinear architectures.

Operationally, we train both predictive architectures on a family of initial states spanning dynamical regimes from chaotic thermalizing behavior to non-ergodic evolution associated with quantum many-body scars. The forecasting task is to predict the future evolution of a set of physical observables, namely local magnetizations and connected two-point correlation functions, given an initial segment of their time series. The performance of the two models is subsequently evaluated on unseen initial conditions and future time windows. We benchmark this framework on a few-qubit PXP chain, for which numerically exact time evolution can be efficiently computed to generate training and test data, enabling a controlled comparison between the two forecasting architectures. Although demonstrated here for the PXP model, the framework is not tied to a specific Hamiltonian and can be extended to other quantum many-body systems, as well as to digital quantum circuits, whenever representative time-series data are available.

\section{The Physical Problem}

We consider a one-dimensional chain of qubits described by the PXP model~\cite{bernien2017probing,serbyn2021quantum, turner2018quantum, liang2025observation}
\begin{equation}
    \hat{H}_{\rm PXP} = \Omega \left(\hat{X}_1 \hat{P}_2 + \sum_{j=2}^{L-1}\hat{P}_{j-1}\hat{X}_j \hat{P}_{j+1} + \hat{P}_{L-1}\hat{X}_L\right),
\end{equation}
where $\hat{Z}_j = \ket{1}_j\bra{1} - \ket{0}_j\bra{0}$ is the $z$-Pauli matrix, $\hat{X}_j = \ket{1}_j\bra{0} + \ket{0}_j\bra{1}$ is the $x$-Pauli matrix, $\hat{P}_j = \frac{1}{2}(\hat{\mathds{1}}_j - \hat{Z}_j)$ is the projector onto the state $\ket{0}_j$, while $\hat{N}_j = \ket{1}_j\bra{1}$. We set $\Omega=1$, so that all physical times discussed in this work are expressed in units of $\Omega^{-1}$.

The PXP model naturally arises in systems of interacting Rydberg atoms operating in the blockade regime~\cite{turner2018quantum, bernien2017probing, lesanovsky2012interacting, serbyn2021quantum, liang2025observation}. Its spectrum is predominantly chaotic, with level statistics and eigenstate properties consistent with the eigenstate thermalization hypothesis (ETH)~\cite{dalessio2016from, deutsch2018eigenstate, srednicki1994chaos, rigol2008themralization}, although it hosts a small set of non-thermal eigenstates forming the quantum many-body scar (QMBS) subspace embedded within an otherwise thermal spectrum~\cite{turner2018quantum}. This coexistence of quantum chaos and atypical eigenstates is reflected in the dynamics. The fully polarized state $\ket{\boldsymbol{0}}=\ket{0000\ldots}$ rapidly explores the accessible Hilbert space and exhibits ergodic behavior consistent with ETH, while the Néel state $\ket{\mathbb{Z}_2}=\ket{1010\ldots}$ shows long-lived revivals due to its overlap with the QMBS subspace~\cite{turner2018quantum}. Interestingly, a richer variety of dynamical behaviors emerges for initial states interpolating between these two limits, including local reminiscent states~\cite{perciavalle2025local}.

The coexistence of distinct dynamical regimes depending on the initial state makes the PXP model both particularly rich and challenging to study, analytically and numerically. In the ergodic regime, the system rapidly generates large amounts of entanglement with a volume-law scaling, which severely limits the efficiency of tensor-network approaches and makes long-time simulations with Matrix Product States (MPS) generally intractable~\cite{cirac2021matrix}. By contrast, non-ergodic initial states can exhibit much slower entanglement growth, allowing for longer accessible simulation times and highlighting the strong dependence of the dynamics on the chosen initial condition.

This work is motivated by these challenges and aims to establish a data-driven framework for learning the dynamics of a class of quantum states from a finite training set, where the class encompasses a variety of dynamical behaviors ranging from chaotic to quantum scarred regimes, thereby enabling prediction of previously unseen dynamical trajectories using ML methods. The general protocol is illustrated in Fig.~\ref{fig:sketch_} and is described in the following.

The class of states on which we train the ML models is given by
\begin{equation}
\ket{\psi_{\lambda,\theta}}=\mathcal{N}(\lambda,\theta)\left[\lambda \ket{\Theta} + (1-\lambda)\ket{\Theta'}\right],
\label{eq:theta_class}
\end{equation}
where $\ket{\Theta}=\ket{\theta,0,\theta,0,\ldots}$ and $\ket{\Theta'}=\ket{0,\theta,0,\theta,\ldots}$ are staggered product states, with $\ket{\theta}=\cos\theta\ket{0}+\sin\theta\ket{1}$, $\theta\in [0,\pi/2]$ and $\lambda\in[0,1]$. Here, $\mathcal{N}(\lambda,\theta)$ denotes a normalization factor. This construction defines a family of states that smoothly interpolates between several physically relevant limits, including Néel and fully polarized configurations, which are recovered for specific parameter choices. In particular, the Néel states $\ket{1010\ldots}$ and $\ket{0101\ldots}$ are obtained for $\theta=\pi/2$ with $\lambda=1$ and $\lambda=0$, respectively, while the fully polarized state $\ket{000\ldots}$ is recovered for $\theta=0$, independently of $\lambda$. Moreover, allowing intermediate values of $\lambda$ introduces entangled initial states, thereby increasing the diversity and complexity of the training set used for the ML models.

Each parameter pair $k=(\lambda_k,\theta_k)$ defines an initial state $\ket{\psi_k}\equiv\ket{\psi_{\lambda,\theta}}$. We consider a set of $N$ initial (input) states, $\{\ket{\psi_k}\}_{k=1}^{N}$, as illustrated in Fig.~\ref{fig:sketch_}(a), generated by random sampling in the $(\lambda,\theta)$ parameter space. For each input state, the corresponding time-evolution trajectory is obtained by evolving the state under the Schrödinger equation, $\ket{\psi_k(t)} = e^{-i\hat H_{\rm PXP} t}\ket{\psi_k}$, where we set $\hbar=1$. To characterize the dynamics, we associate with each trajectory a time-dependent feature vector $\mathbf{x}_k(t)$, constructed from physical observables, specifically local magnetizations and connected two-point correlations.

Time is discretized into $T$ equally spaced steps with spacing $\Delta t$, such that the physical times are $t_i=(i-1)\Delta t$ for $i=1,\ldots,T$. At each time $t_i$, we define the feature vector $\mathbf{x}_{k,i}\equiv \mathbf{x}_k(t_i)$, which encodes the state of the system at that time via the chosen set of observables:
\begin{equation}
\mathbf{x}_{k,i}
=
\Big[
\{(m_j^z(t_i))_k\}_{j=1}^{L},
\;
\{(c_{jl}^{zz}(t_i))_k\}_{1\le j<l\le L}
\Big]\in \mathbb{R}^F,
\end{equation}
namely as the concatenation of all single-site magnetizations and connected pairwise correlators. Here,
$(m_j^z(t))_k=\langle\psi_k(t)|\hat Z_j|\psi_k(t)\rangle$ and
$(c_{jl}^{zz}(t))_k=\langle\psi_k(t)|\hat Z_j\hat Z_l|\psi_k(t)\rangle
-\langle\psi_k(t)|\hat Z_j|\psi_k(t)\rangle
\langle\psi_k(t)|\hat Z_l|\psi_k(t)\rangle$. The total number of features is $F=L(L+1)/2$, comprising $L$ local magnetizations and $L(L-1)/2$ connected two-body correlators. In this work, we consider a system composed if $L=10$ qubits which corresponds to a number of features $F=55$.
%

\section{Methods}

\subsection{Formal Definitions}
Each trajectory $k$ is represented by a sequence of $T$ feature
vectors $\mathbf{x}_{k,i}\in\mathbb{R}^F$. We divide each trajectory into
an input sequence containing the first $\rho<T$ time steps,
\begin{equation}
    X_k^{\mathrm{in}}
    =
    \left(
    \mathbf{x}_{k,1},\ldots,\mathbf{x}_{k,\rho}
    \right)
    \in\mathbb{R}^{\rho\times F},
\end{equation}
and a target sequence containing the subsequent $\tau=T-\rho$ time steps,
\begin{equation}
    X_k^{\mathrm{out}}
    =
    \left(
    \mathbf{x}_{k,\rho+1},\ldots,\mathbf{x}_{k,T}
    \right)
    \in\mathbb{R}^{\tau\times F}.
\end{equation}

The forecasting dataset is therefore defined as
\begin{equation}
    \mathcal{D}
    =
    \left\{
    \left(X_k^{\mathrm{in}},X_k^{\mathrm{out}}\right)
    \right\}_{k=1}^{N}.
\end{equation}
The objective is to learn a mapping
\begin{equation}
    f:\mathbb{R}^{\rho\times F}
    \rightarrow
    \mathbb{R}^{\tau\times F},
\end{equation}
such that
\begin{equation}
    \tilde{X}_k^{\mathrm{out}}
    =
    f\left(X_k^{\mathrm{in}}\right)
\end{equation}
approximates the future trajectory $X_k^{\mathrm{out}}$ [Fig.~\ref{fig:sketch_}(b),(c)].

In practice, $N_{\mathrm{train}}<N$ trajectories are used for training,
while the remaining
$N_{\mathrm{test}}=N-N_{\mathrm{train}}$ trajectories are reserved for
evaluation. We consider a DLinear model~\cite{zeng2023transformers} and a
Transformer architecture~\cite{vaswani2017attention} for learning this
mapping.

\subsection{Models}
We consider two complementary neural architectures for time-series forecasting: DLinear~\cite{zeng2023transformers, li2026revisiting} and the Transformer~\cite{vaswani2017attention, sutskever2014sequence}. The two models differ primarily in how they generate future predictions. DLinear is a direct multi-step forecasting architecture that predicts the entire future horizon in a single forward pass, whereas the Transformer is employed in an autoregressive framework, generating future states sequentially. This distinction allows us to compare direct and autoregressive forecasting strategies under a common training objective.

Both models are trained using the same \emph{informed} composite loss, which incorporates domain-relevant information beyond pointwise prediction accuracy. In addition to penalizing errors in the predicted states, the loss promotes temporal consistency by penalizing discrepancies in the evolution of the predicted trajectory, including differences in temporal increments and inconsistencies in their direction. Consequently, the training objective encourages both architectures not only to reconstruct the target states accurately, but also to preserve relevant dynamical characteristics of the underlying system. 
\subsubsection{DLinear}

DLinear is a simple yet effective linear architecture for time-series forecasting, introduced by Zeng \emph{et al.}~\cite{zeng2023transformers}. Unlike autoregressive models, DLinear performs direct multi-step forecasting by predicting the entire future horizon in a single forward pass. This design avoids the recursive generation of future states and, consequently, does not suffer from error accumulation associated with autoregressive prediction or the exposure bias arising from the discrepancy between training and inference inputs.

The model first decomposes the input sequence into trend and seasonal components through a moving-average operation. The trend component captures the slowly varying, long-term behavior of the signal, while the seasonal component represents recurring variations around this trend. Two independent linear layers are then applied to the decomposed components, and their outputs are combined to produce the final prediction. Despite its simplicity, DLinear has been shown to achieve competitive or superior performance compared to more complex Transformer-based architectures in several long-term forecasting benchmarks~\cite{zeng2023transformers}.

\subsubsection{Transformer}

Transformers are DL architectures originally introduced for sequence modeling in natural language processing. Unlike recurrent neural networks, they process the entire input sequence simultaneously and capture long-range dependencies through the self-attention mechanism, which enables the model to dynamically weigh the relevance of different elements of the input sequence when generating predictions. Since their introduction, Transformers have become widely used for sequential learning tasks, including time-series forecasting and scientific ML. The Transformer architecture was first proposed by Vaswani \emph{et al.}~\cite{vaswani2017attention}.

In this work, we employ the Transformer in an autoregressive forecasting framework. During training, the model receives as input a sequence of $\rho$ consecutive time steps and is trained to predict only the subsequent step. During inference, the predicted state is appended to the input sequence, replacing the oldest time step through a sliding-window mechanism, and the updated sequence is fed back to the model to predict the next state. This iterative procedure is repeated until the desired prediction horizon $\tau$ is reached.

More formally, the Transformer models the probability of the future sequence as
\begin{equation}
\begin{aligned}
&p\!\left(
\mathbf{x}_{\rho+1},\ldots,\mathbf{x}_{\rho+\tau}
\mid
\mathbf{x}_1,\ldots,\mathbf{x}_{\rho}
\right) \\
&\quad =
\prod_{i=\rho+1}^{\rho + \tau}
p\!\left(
\mathbf{x}_{i}
\mid
\mathbf{x}_1,\ldots,\mathbf{x}_{i-1}
\right).
\end{aligned}
\end{equation}

Despite its ability to capture long-range dependencies, autoregressive forecasting suffers from two well-known limitations. First, prediction errors can accumulate over time, since each predicted state is recursively used to generate subsequent predictions, potentially leading to progressively larger deviations over long horizons. Second, training is typically performed using teacher forcing, where the model conditions on ground-truth states, whereas during inference it relies on its own predictions. This discrepancy between the training and inference conditions is known as \emph{exposure bias}. In contrast to DLinear, which predicts the entire horizon directly, the autoregressive Transformer is therefore subject to these recursive effects.

\subsection{Numerical Experiments}

We benchmark the Transformer and DLinear on a dataset comprising
$N=500$ trajectories generated from initial states sampled in the
$(\lambda,\theta)$ parameter space. Each trajectory contains $T=800$
uniformly spaced time steps, with $\Delta t=0.02$, and is represented by
$F=55$ physical observables corresponding to the $L=10$ local
magnetizations and the $L(L-1)/2=45$ connected two-point correlators.
Consequently, the complete dataset can be represented as a tensor
$X\in\mathbb{R}^{500\times800\times55}$.

The dataset is split at the trajectory level before preprocessing. The first
$N_{\rm test}=100$ trajectories are held out for testing, while the remaining
$N_{\rm train}=400$ trajectories are used for training and validation. This
ensures that different time steps belonging to the same physical trajectory
never appear in both the training and test sets. A feature-wise Min--Max
transformation to the interval $[0,1]$ is fitted exclusively on the training
trajectories and subsequently applied, without refitting, to the test set.
Predictions are transformed back to the original physical scale before the
evaluation metrics are computed.

For every trajectory, the models receive the first $\rho=500$ time steps as
input and predict the subsequent $\tau=300$ steps. Since $\Delta t=0.02$, the
input window covers the physical-time interval $t\in[0,10)$, whereas the
forecasting horizon covers $t\in[10,16)$. Both architectures therefore address
the same forecasting problem,
\begin{equation}
    f_{\mathcal{M}}:
    \mathbb{R}^{500\times55}
    \longrightarrow
    \mathbb{R}^{300\times55},
\end{equation}
where $\mathcal{M}\in\{\mathtt{TR},\mathtt{DL}\}$ denotes, respectively, the
 Transformer and DLinear. The two models differ only in how this mapping
is implemented. DLinear predicts the complete future trajectory in a single
forward pass, whereas the  Transformer is trained using teacher forcing
and generates the forecast autoregressively at inference time.

\subsubsection{Training \& Instantiations}

To provide a controlled comparison, both models are trained using the same
temporally informed objective. For a mini-batch containing $B$ trajectories,
let $\textrm{x}_{k,i,f}$ and $\tilde{\textrm{x}}^{(\mathcal{M})}_{k,i,f}$ denote the true and
predicted values of feature $f$ at forecasting step $i$, respectively. The
pointwise reconstruction term is
\begin{equation}
    \mathcal{L}_{\mathrm{pos}}^{(\mathcal{M})}
    =
    \frac{1}{B\tau F}
    \sum_{k=1}^{B}
    \sum_{i=\rho + 1}^{\rho+\tau}
    \sum_{f=1}^{F}
    \left(
        \textrm{x}_{k,i,f}
        -
        \tilde{\textrm{x}}^{(\mathcal{M})}_{k,i,f}
    \right)^2.
\end{equation}
To promote agreement between the local temporal variations of the predicted
and exact trajectories, we define
\begin{equation}
    \Delta \textrm{x}_{k,i,f}
    =
    \textrm{x}_{k,i,f}-\textrm{x}_{k,i-1,f},
    \qquad
    \Delta\tilde{\textrm{x}}^{(\mathcal{M})}_{k,i,f}
    =
    \tilde{\textrm{x}}^{(\mathcal{M})}_{k,i,f}
    -
    \tilde{\textrm{x}}^{(\mathcal{M})}_{k,i-1,f},
\end{equation}
for $i=\rho+2,\ldots,\rho+\tau$. The derivative-consistency term is then
\begin{equation}
    \mathcal{L}_{\mathrm{dir}}^{(\mathcal{M})}
    =
    \frac{1}{B(\tau-1)F}
    \sum_{k=1}^{B}
    \sum_{i=\rho + 2}^{\rho+\tau}
    \sum_{f=1}^{F}
    \left(
        \Delta \textrm{x}_{k,i,f}
        -
        \Delta\tilde{\textrm{x}}^{(\mathcal{M})}_{k,i,f}
    \right)^2,
\end{equation}
while directional disagreements are penalized through
\begin{equation}
    \mathcal{L}_{\mathrm{sign}}^{(\mathcal{M})}
    =
    \frac{1}{B(\tau-1)F}
    \sum_{k=1}^{B}
    \sum_{i=\rho + 2}^{\rho+\tau}
    \sum_{f=1}^{F}
    \operatorname{ReLU}
    \left(
        -\Delta \textrm{x}_{k,i,f}
        \Delta\tilde{\textrm{x}}^{(\mathcal{M})}_{k,i,f}
    \right).
\end{equation}
The complete training objective is
\begin{equation}
    \mathcal{L}^{(\mathcal{M})}
    =
    \mathcal{L}_{\mathrm{pos}}^{(\mathcal{M})}
    +
    \alpha\mathcal{L}_{\mathrm{dir}}^{(\mathcal{M})}
    +
    \beta\mathcal{L}_{\mathrm{sign}}^{(\mathcal{M})},
    \qquad
    \alpha=0.5,\quad\beta=2.0.
\end{equation}
This loss does not explicitly impose the equations of motion. Instead, it
introduces a soft temporal inductive bias that encourages the models to reproduce
the values, local increments, and directions of variation of the physical
observables.

The  Transformer is instantiated with model dimension
$d_{\mathrm{model}}=256$, seven encoder--decoder layers, two attention heads,
and a feed-forward dimension $d_{\mathrm{ff}}=256$. These architectural
hyperparameters were selected through Hyperband search on the
training set using validation loss as the selection criterion. The model is
optimized with AdamW using a learning rate of $3\times10^{-4}$, weight decay
$10^{-3}$, gradient clipping with norm $1.0$, and a batch size of $4$. Training
is performed for at most $100$ epochs, with a $30\%$ validation split, a
\texttt{ReduceLROnPlateau} scheduler, and early stopping with patience $8$.

During training, the Transformer decoder receives the shifted ground-truth
target sequence,
\begin{equation}
    D_k=
    \left(
        \boldsymbol{0},
        \mathbf{x}_{k,\rho+1},
        \ldots,
        \mathbf{x}_{k,\rho+\tau-1}
    \right),
\end{equation}
and a causal mask prevents access to subsequent target steps. At inference time,
ground-truth future values are unavailable; therefore, the decoder is initialized
with a zero vector and filled recursively with its own previous predictions.
The resulting model is referred to as the  Transformer because it is
trained with the temporally informed composite loss defined above.

For DLinear, the moving-average window is set to $w=7$, corresponding to
$7\Delta t=0.14$ physical-time units, and the selected $L^2$ regularization
coefficient is $10^{-5}$. These values were obtained through a Bayesian
hyperparameter search on the training set. DLinear is optimized with AdamW using
a learning rate of $10^{-3}$, weight decay $10^{-3}$, and a batch size of $64$.
Training is performed for at most $250$ epochs with a $30\%$ validation split,
a \texttt{ReduceLROnPlateau} scheduler, and early stopping with patience $10$.
Unlike the  Transformer, DLinear directly produces all $\tau=300$
future steps in a single forward pass and therefore does not use teacher forcing
or autoregressive feedback.

The preprocessing procedure, trajectory-level split, input window, forecasting
horizon, loss weights, and evaluation protocol are identical for the two
architectures. Model-specific optimization settings are selected independently
to account for their substantially different computational and memory
requirements.

\subsubsection{Evaluation Metrics}
The forecasting performance of the models is evaluated on the
$N_{\rm test}$ trajectories that are not used during training. We consider
two complementary metrics: the mean absolute error (MAE), which quantifies
the magnitude of the prediction error, and the directional accuracy (DA),
which measures the ability of the model to reproduce the direction of the
temporal evolution of each observable.

Let $\textrm{x}_{k,i,f}$ and $\tilde{\textrm{x}}_{k,i,f}$ denote, respectively, the true and
predicted values of feature $f$ at the $i$-th step of the prediction horizon
for test trajectory $k$. The mean absolute error is defined as

\begin{equation}
\mathrm{MAE}
=
\frac{1}{N_{\rm test}\tau F}
\sum_{k=1}^{N_{\rm test}}
\sum_{i=\rho +1}^{\rho + \tau}
\sum_{f=1}^{F}
\left|
\tilde{\textrm{x}}_{k,i,f} - \textrm{x}_{k,i,f}
\right|.
\label{eq:mae}
\end{equation}

The MAE therefore measures the average absolute deviation between the
predicted and exact trajectories over all test samples, prediction steps,
and physical observables. Lower values of the MAE correspond to more
accurate predictions.

To complement this amplitude-based measure, we evaluate the directional
accuracy (DA), which quantifies whether the model correctly reproduces the
sign of the local temporal variation of each observable, independently of
how well it reproduces its absolute magnitude. To make this precise, we
first define the true and predicted increments of each feature between
consecutive time steps,
\begin{equation}
\Delta \textrm{x}_{k,i,f}
=
\textrm{x}_{k,i,f}-\textrm{x}_{k,i-1,f},
\qquad
\Delta \tilde{\textrm{x}}_{k,i,f}
=
\tilde{\textrm{x}}_{k,i,f}-\tilde{\textrm{x}}_{k,i-1,f}.
\label{eq:increments}
\end{equation}
The DA is then defined as the percentage of (trajectory, time step, feature)
triples for which the predicted and true increments share the same sign,

\begin{equation}
\begin{aligned}
\mathrm{DA}
&=
\frac{100}{N_{\rm test}\tau F}
\sum_{k=1}^{N_{\rm test}}
\sum_{i=\rho +1}^{\rho + \tau}
\sum_{f=1}^{F}
\\
&\quad \times
\mathbf{1}
\left[
\operatorname{sgn}(\Delta\tilde{\textrm{x}}_{k,i,f})
=
\operatorname{sgn}(\Delta \textrm{x}_{k,i,f})
\right],
\end{aligned}
\label{eq:da}
\end{equation}

where $\mathbf{1}[\cdot]$ is the indicator function, equal to one when its
argument is true and zero otherwise. For the first predicted time step
($i=\rho + 1$), the last observed value of the input sequence is used as the
reference, i.e., $\tilde{\textrm{x}}_{k,\rho,f}=\textrm{x}_{k,\rho,f}$. The DA is
reported as a percentage, with $100\%$ corresponding to a prediction that
reproduces the correct direction of change for every feature and time step.

\begin{figure*}[!t]
    \centering
    \includegraphics[width=\linewidth]{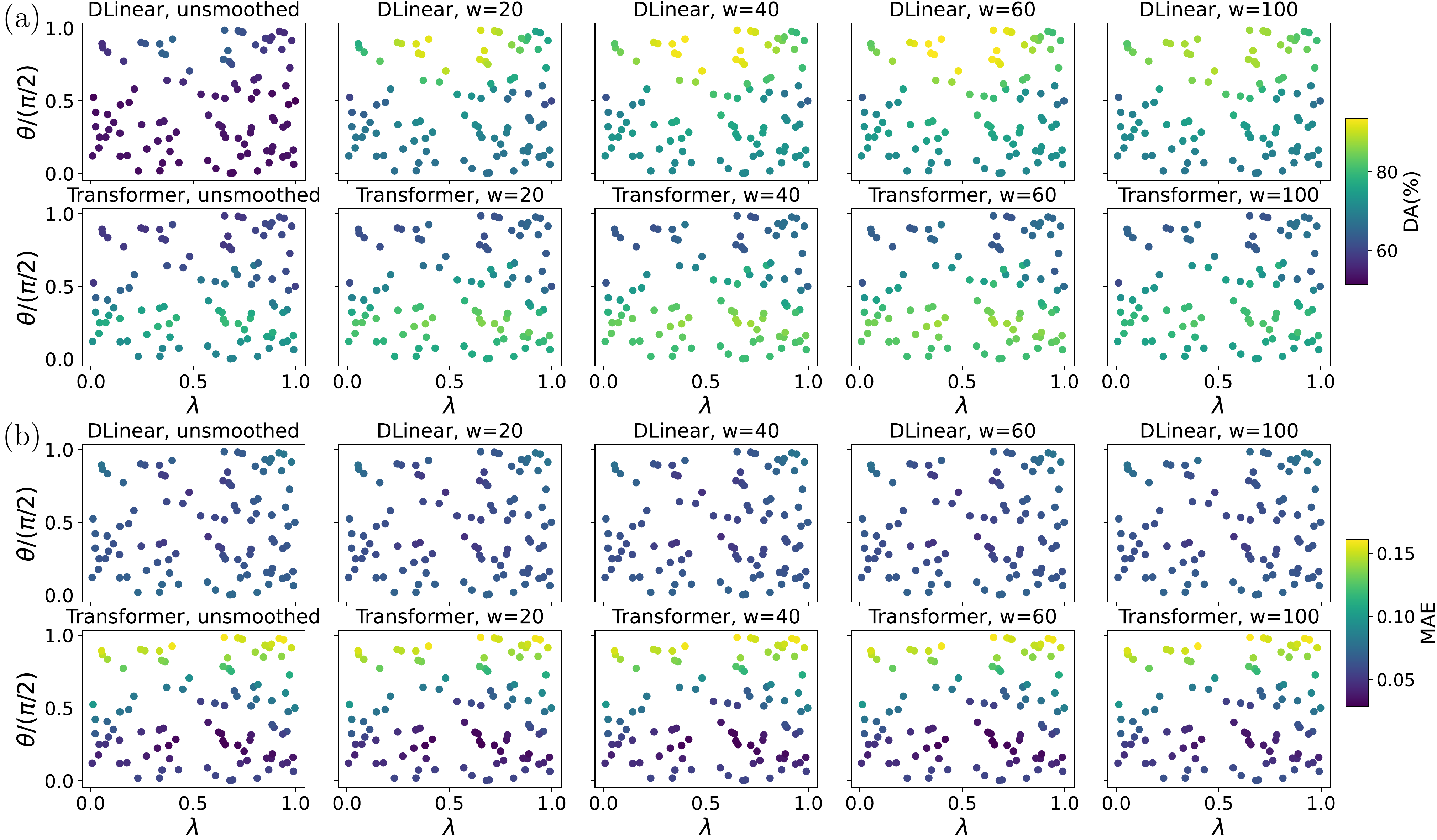}
    \caption{Performance of DLinear and the Transformer across the initial-state parameter
space $(\lambda,\theta)$. Panel (a) shows the directional accuracy (DA), while
panel (b) shows the mean absolute error (MAE). In each panel, the upper row
corresponds to DLinear and the lower row to the Transformer. The leftmost
column reports the metrics computed from the raw predictions, while the
remaining columns show the corresponding results after smoothing the predicted
trajectories using windows $w\in\{20,40,60,100\}$. In all cases, the metrics
are evaluated with respect to the ground-truth trajectories.}
    \label{fig:DA_MAE}
\end{figure*}
The two metrics provide complementary information. While the MAE evaluates
how closely the predicted trajectories reproduce the numerical values of
the physical observables, the DA evaluates whether the model captures their
local temporal trends. More generally, MAE and DA are not equivalent
measures of forecasting quality, and a model can score well on one while
scoring poorly on the other. Oscillatory dynamics provide an example of
this: a model may correctly reproduce the oscillation frequency, rising and
falling at the right times and therefore achieving a high DA, while getting
the oscillation amplitude wrong, which inflates the MAE. Conversely, a 
related situation arises when the true dynamics consists of a single,
clean oscillation, and the model correctly reproduces its frequency and
amplitude, but its prediction is modulated by an additional, spurious
high-frequency oscillation of small amplitude superimposed on top of the
correctly predicted signal. Because this spurious modulation has small
amplitude, the prediction remains numerically close to the true trajectory
at every step, and the MAE stays low. However, at the time steps where the
spurious oscillation locally reverses the sign of the predicted increment,
the predicted direction of motion disagrees with the true one, even though
the true dynamics is not oscillating at that frequency at all; this
lowers the DA despite the small MAE.

The DA is therefore a purely qualitative metric: it does not depend on the
size of the error, only on its sign, and it treats a barely-missed
direction the same as a badly-missed one, and a prediction that is off by
an arbitrarily large amount the same as one that is off by an
infinitesimal one, as long as the sign of the step-to-step change is
correct. This makes the DA comparatively insensitive to outliers and to the
overall scale of the observable, and therefore particularly informative
when comparing performance across different features or different
observables of the system, whose magnitudes and dynamical ranges need not
coincide, a comparison in which the MAE alone would be dominated by the
observables with the largest dynamical range. At the same time, since the
DA discards all amplitude information, it cannot by itself distinguish a
model that tracks the dynamics closely from one that reproduces only its
coarse qualitative pattern; a high DA is therefore a necessary, but not
sufficient, condition for an accurate reconstruction of the dynamics, and
must be interpreted jointly with the MAE.
\section{Results}
\begin{figure}[t]
    \centering
    \includegraphics[width=\linewidth]{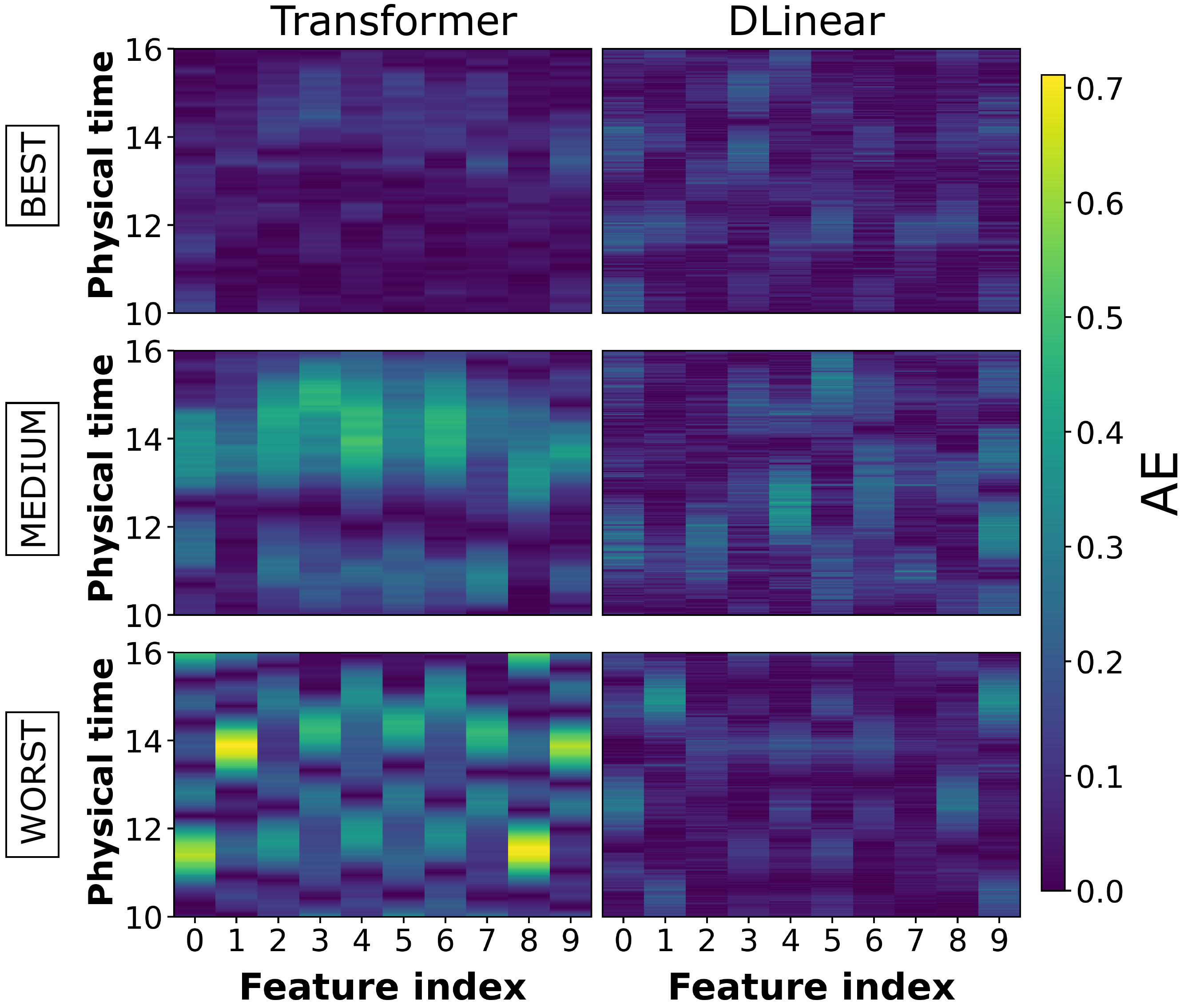}
    \caption{Absolute error (AE) over the forecast horizon for the first ten features,
corresponding to the local magnetizations. The vertical axis represents
physical time and the horizontal axis the feature index. The left and right
columns show the Transformer and DLinear predictions, respectively. The three
rows correspond to representative test trajectories selected as the best,
typical, and worst cases using a joint score for the two models. For each test
trajectory, the MAE is first computed by averaging over the forecast horizon
and all features, and the resulting values for the Transformer and DLinear are
then averaged to define the joint score. The best and worst cases correspond
to the lowest and highest joint MAE, respectively, while the typical case is
the trajectory whose joint MAE is closest to the test-set average. The color
scale indicates the magnitude of the absolute error and is shared across all
panels.}
    \label{fig:global_absolute_prediction_error_heatmaps}
\end{figure}

\paragraph{Trajectory-resolved forecasting performance}
We first analyze the predictive performance of the two models across the
parameter space $(\lambda,\theta)$ defining the initial states. Since each
test trajectory $k$ corresponds to a distinct initial condition
$(\lambda_k,\theta_k)$, we evaluate the metrics introduced in
Eqs.~\eqref{eq:mae} and~\eqref{eq:da} separately for each trajectory,
averaging only over the prediction horizon and the features:
\begin{align}
\mathrm{MAE}_k
&=
\frac{1}{\tau F}
\sum_{i=\rho+1}^{\rho+\tau}
\sum_{f=1}^{F}
\left|
\tilde{\textrm{x}}_{k,i,f}-\textrm{x}_{k,i,f}
\right|,
\\
\mathrm{DA}_k
&=
\frac{100}{\tau F}
\sum_{i=\rho+1}^{\rho+\tau}
\sum_{f=1}^{F}
\mathbf{1}
\left[
\operatorname{sgn}(\Delta\tilde{\textrm{x}}_{k,i,f})
=
\operatorname{sgn}(\Delta\textrm{x}_{k,i,f})
\right].
\label{eq:mae_da_k}
\end{align}

The predicted trajectories can exhibit small-amplitude, high-frequency
oscillations that strongly affect the directional accuracy, despite having
only a minor impact on the overall agreement with the target trajectories.
To disentangle these rapid fluctuations from the underlying temporal trend,
in Fig.~\ref{fig:DA_MAE} we also report metrics evaluated after smoothing the
predicted trajectories. We employ a Savitzky--Golay
filter~\cite{savitzky1964smoothing}, which locally approximates the signal
with a second-order polynomial over a sliding window of length $w$.
Increasing $w$ progressively suppresses short-timescale variations, while
excessively large windows may also remove genuine dynamical features.
Importantly, smoothing is used only as a post-processing tool for this
analysis and does not affect the training procedure or the raw model
predictions.

Figure~\ref{fig:DA_MAE}(a) shows the directional accuracy for DLinear
(upper panels) and the Transformer (lower panels), starting from the raw
predictions and progressively increasing the smoothing window $w$ from left
to right. For the raw DLinear predictions, $\mathrm{DA}_k$ is approximately
uniform over the initial-state parameter space and remains close to $50\%$.
In contrast, the Transformer displays a pronounced dependence on $\theta$,
with substantially higher directional accuracy for
$\theta\lesssim\pi/4$ and lower values as the initial states approach the
N\'eel limit, while showing little dependence on $\lambda$.

The behavior of DLinear changes substantially when the high-frequency
components of the predicted trajectories are filtered out. As $w$ increases,
the directional accuracy rises well above $50\%$ over a broad region of
parameter space, reaching values above $90\%$ for intermediate smoothing
windows, $w\in[20,60]$, in a large portion of the parameter space.
This strong improvement indicates that the low DA of the raw DLinear
predictions originates predominantly from small-amplitude, high-frequency
oscillations rather than from an incorrect description of the underlying
temporal trend. For $w=100$, the DA decreases again, indicating that the
smoothing window has become sufficiently large to suppress relevant
variations of the dynamics together with the high-frequency components.

The Transformer is considerably less sensitive to smoothing. In particular,
the separation between the regions $\theta\lesssim\pi/4$ and
$\theta\gtrsim\pi/4$ persists for all the smoothing windows considered.
Although intermediate values of $w$ improve the directional accuracy to some
extent, smoothing does not remove the loss of performance for
$\theta\gtrsim\pi/4$. This indicates that, in this region, the Transformer
does not merely superimpose high-frequency fluctuations on an otherwise
accurate forecast, but increasingly fails to reproduce the temporal trend of
the target dynamics.

A complementary picture is provided by the mean absolute error,
shown in Fig.~\ref{fig:DA_MAE}(b). Unlike DA, which is highly sensitive to
the sign of consecutive variations, MAE measures the average pointwise
deviation between predicted and target trajectories and is therefore much
less affected by small-amplitude, high-frequency oscillations. Accordingly,
the MAE depends only weakly on the smoothing procedure. For DLinear, it
remains low and approximately uniform throughout the $(\lambda,\theta)$
parameter space. The Transformer instead exhibits a clear dependence on
$\theta$: its MAE is comparable to that of DLinear for
$\theta\lesssim\pi/4$, but increases substantially for
$\theta\gtrsim\pi/4$, with little dependence on $\lambda$.

Taken together, the two metrics reveal qualitatively different behaviors of
the forecasting models. DLinear maintains a low MAE throughout the entire
family of initial states, while the apparent loss of directional accuracy in
the raw predictions can largely be attributed to small-amplitude,
high-frequency oscillations. Once these components are filtered out, the
underlying trend is recovered with high directional accuracy. The
Transformer, by contrast, performs well in the regime
$\theta\lesssim\pi/4$ but progressively loses predictive accuracy as the
initial states approach the quantum-scarred regime. Its performance is
therefore strongly dependent on the dynamical regime, whereas DLinear
provides a considerably more uniform forecast across the initial-state
parameter space.

\paragraph{Feature- and time-resolved prediction errors}
To examine the forecasting performance at the level of individual
trajectories and observables, Fig.~\ref{fig:global_absolute_prediction_error_heatmaps}
shows the absolute error (AE),
\begin{equation}
    \mathrm{AE}_{k,i,f}
    =
    \left|
    \tilde{\textrm{x}}_{k,i,f}
    -
    \textrm{x}_{k,i,f}
    \right|,
\end{equation}
where $\tilde{\textrm{x}}_{k,i,f}$ and $\textrm{x}_{k,i,f}$ denote the
predicted and target values of feature $f$ at time step $i$ of test
trajectory $k$. The AE is the pointwise counterpart of the MAE introduced
above, which is obtained by averaging over trajectories, time steps, and
features.

We consider three representative test trajectories, selected using the
performance of both models on the same initial conditions. For each
trajectory, we compute the MAE of each model over the prediction horizon and
all features, and average the two values to define a joint score. The
trajectories with the lowest and highest joint MAE are selected as the best
and worst cases, respectively, while the trajectory whose joint MAE is
closest to the test-set average is selected as the typical case. The rows in
Fig.~\ref{fig:global_absolute_prediction_error_heatmaps} correspond to these
three trajectories, ordered from best to worst, while the columns compare the
Transformer and DLinear. The horizontal axis shows the first ten features,
corresponding to the local magnetizations, and the vertical axis spans the
prediction interval $t\in[10,16)$, following the input interval
$t\in[0,10)$. A common color scale across all panels reports the magnitude of
the absolute error.

The heatmaps reveal qualitatively different error patterns for the two
models. For DLinear, the errors remain comparatively small across the three
representative trajectories. Their rapid variations in time help explain the
low directional accuracy of the raw predictions observed in
Fig.~\ref{fig:DA_MAE}: even when the predicted trajectory remains close to
the target in magnitude, small-amplitude, high-frequency deviations can
repeatedly change the sign of consecutive variations and thereby reduce the
DA. The strong increase in DA after smoothing is consistent with the
suppression of these short-timescale deviations while leaving the overall
trajectory largely unchanged.

The Transformer exhibits a stronger dependence on the initial condition.
For trajectories where its forecasting performance is favorable, the errors
remain relatively small throughout the prediction horizon. In the less
favorable regime identified in Fig.~\ref{fig:DA_MAE}, however, larger errors
persist over extended portions of the prediction window and across several
observables. This behavior is most evident for the worst-case trajectory,
where the absolute error of the Transformer locally approaches $0.7$, while
DLinear retains substantially smaller errors over most of the prediction
window.

These feature- and time-resolved errors complement the aggregate results of
Fig.~\ref{fig:DA_MAE}. For DLinear, the low MAE together with the strong
effect of smoothing on DA indicates that its main deviations are
small-amplitude and rapidly varying. For the Transformer, the deterioration
observed in the unfavorable parameter regime instead involves larger and
more persistent errors, consistent with a genuine loss of forecasting
accuracy rather than a sensitivity of the DA metric to high-frequency
fluctuations. The local error analysis therefore reinforces the different
behavior of the two models identified from the trajectory-resolved metrics.


\begin{figure}[t]
    \centering
    \includegraphics[width=\linewidth]{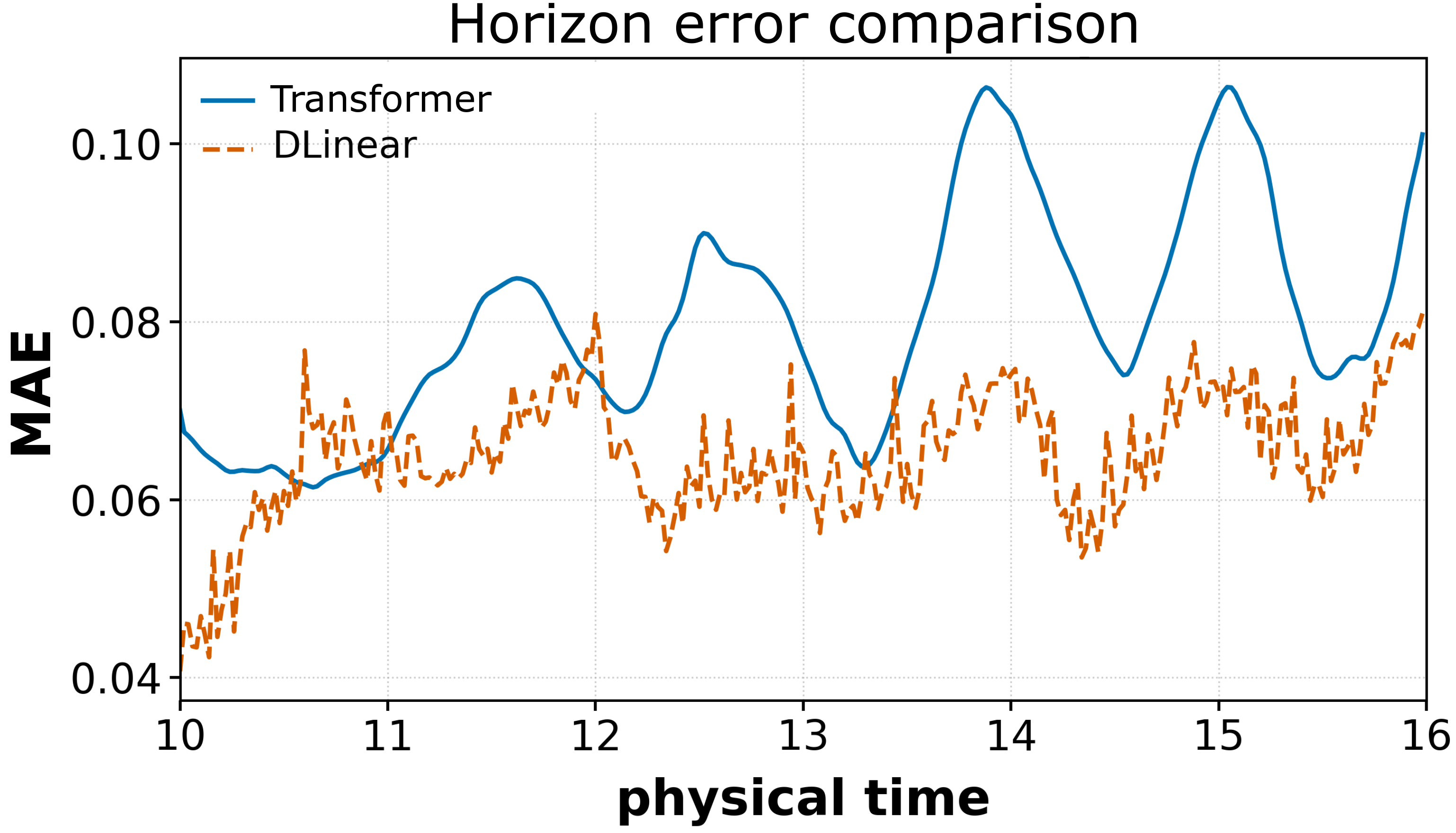}
    \caption{Mean absolute error $\mathrm{MAE}_i$ as a function of physical time over the
forecast horizon $t\in[10,16)$, obtained by averaging over all test trajectories
and features at each prediction time. The solid and dashed lines correspond to
the Transformer and DLinear, respectively. }
    \label{fig:global_horizon_error_comparison}
\end{figure}

\paragraph{Error propagation over the forecast horizon}
We next examine how the prediction error evolves over the forecast horizon.
Figure~\ref{fig:global_horizon_error_comparison} shows the MAE resolved as a
function of the prediction time,
\begin{equation}
\mathrm{MAE}_i
=
\frac{1}{N_{\rm test}F}
\sum_{k=1}^{N_{\rm test}}
\sum_{f=1}^{F}
\left|
\tilde{\textrm{x}}_{k,i,f}
-
\textrm{x}_{k,i,f}
\right|,
\end{equation}
where $i=\rho+1,\ldots,\rho+\tau$ and the corresponding physical time is
$t_i=(i-1)\Delta t$.

For DLinear, the MAE initially increases and subsequently settles into a
bounded, approximately stationary regime, fluctuating around a nearly
constant value without a clear systematic increase over the remainder of the
forecast horizon. The residual fluctuations occur on relatively short
timescales and remain small in amplitude, so that the prediction error does
not accumulate appreciably as the forecast progresses. This behavior is
consistent with the trajectory-resolved results of
Fig.~\ref{fig:DA_MAE}, where DLinear exhibits a low and comparatively uniform
MAE across the initial-state parameter space.

The Transformer displays a qualitatively different behavior. Its MAE
exhibits pronounced oscillations on longer timescales, together with an
overall increase in the magnitude of the error over the forecast horizon.
In particular, the error peaks become progressively larger at later
prediction times, resulting in a substantially stronger error growth than
for DLinear. This behavior is consistent with the strong dependence of the
Transformer performance on the initial state observed in
Fig.~\ref{fig:DA_MAE}. The larger errors associated with the region
approaching the N\'eel limit, $\theta\gtrsim\pi/4$, contribute substantially
to the trajectory-averaged MAE and therefore to its larger overall magnitude.

The time-resolved analysis thus provides a complementary distinction between
the two forecasting models. While the DLinear error remains bounded after an
initial transient, the Transformer exhibits larger temporal variations and
a stronger growth of the prediction error over the forecast horizon.


\begin{figure}[!t]
    \centering
    \includegraphics[width=\linewidth]{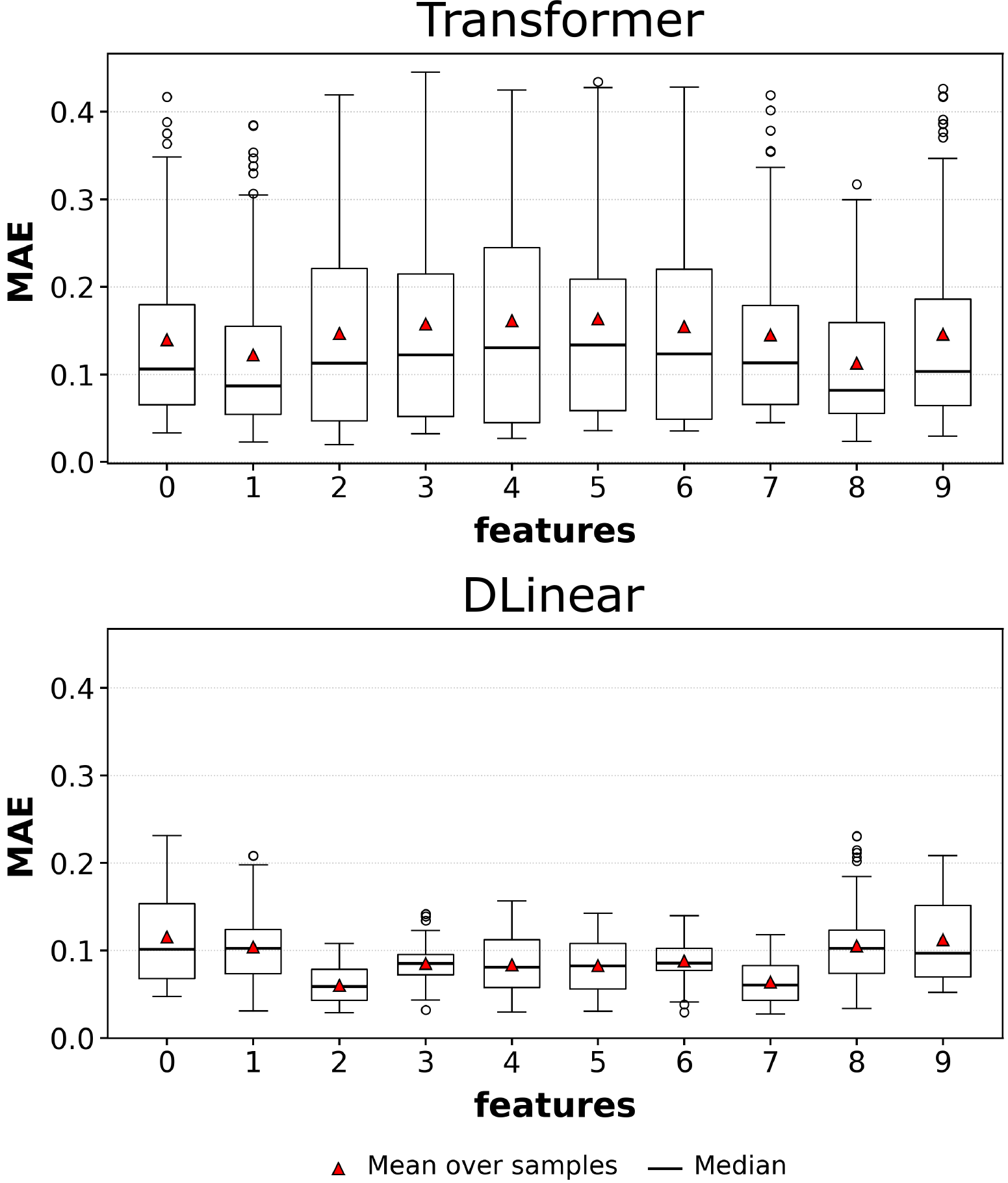}
    \caption{Distribution of the feature-resolved mean absolute error,
$\mathrm{MAE}_{k,f}$, over the test trajectories for the first ten features,
corresponding to the local magnetizations. The upper and lower panels show the
Transformer and DLinear, respectively. For each feature, the boxplot represents
the distribution of $\mathrm{MAE}_{k,f}$ across the test trajectories. Black
lines indicate the median, while red triangles indicate the mean.}
    \label{fig:dlinear_feature_boxplot_mae}
\end{figure}

\paragraph{Feature-resolved prediction errors}
Finally, we examine how the prediction error depends on the individual
observables, focusing on the local magnetizations. For each test trajectory
$k$ and feature $f$, we define the feature-resolved MAE by averaging only
over the prediction horizon,
\begin{equation}
\mathrm{MAE}_{k,f}
=
\frac{1}{\tau}
\sum_{i=\rho+1}^{\rho+\tau}
\left|
\tilde{\textrm{x}}_{k,i,f}
-
\textrm{x}_{k,i,f}
\right|.
\label{eq:mae_kf}
\end{equation}
For each feature $f$, this yields a distribution of $N_{\rm test}$ error
values, one for each test trajectory. Figure~\ref{fig:dlinear_feature_boxplot_mae}
shows these distributions for both forecasting models. Each box represents
the interquartile range, with the horizontal line indicating the median,
while the whiskers extend to the most extreme values within
$1.5\,\mathrm{IQR}$ and points outside this range are shown individually as
outliers. The mean over the test trajectories,
$N_{\rm test}^{-1}\sum_k \mathrm{MAE}_{k,f}$, is indicated by a red triangle.

Features $0,\ldots,9$ correspond to the local magnetizations at sites
$j=1,\ldots,L$ along the chain. Both models display a relatively regular
behavior across sites, although with different spatial profiles. For the
Transformer, the median error is approximately uniform along the chain, while
the distribution broadens around the central sites, where the upper
quartiles and whiskers extend toward larger errors. DLinear exhibits the
opposite tendency: the errors and their dispersion are smaller near the
center of the chain and increase toward the boundaries.

A more pronounced difference between the two models emerges from the
distribution of errors over the test trajectories. DLinear has a lower mean
$\mathrm{MAE}_{k,f}$ for every feature and generally exhibits narrower
distributions. Moreover, its mean and median remain close across the chain,
indicating that the average error is representative of the typical
trajectory. For the Transformer, instead, the mean lies systematically above
the median, revealing a right-skewed distribution with a tail toward larger
errors. At the level of the median, the Transformer is competitive with
DLinear for several features and is even slightly more accurate at some edge
sites, such as features $1$ and $8$. Its larger mean error therefore does not
reflect a uniform degradation across all trajectories, but is driven by a
subset of initial conditions for which the prediction error becomes
substantially larger.

This picture is consistent with the trajectory-resolved analysis of
Fig.~\ref{fig:DA_MAE}, where the Transformer performance depends strongly on
the initial-state parameter $\theta$, while DLinear remains comparatively
uniform across the parameter space. The upper tail of the Transformer error
distributions is consistent with the large errors observed for trajectories
approaching the N\'eel regime, whereas trajectories in the region where the
Transformer performs well contribute to the lower-error part of the
distribution. The feature-resolved analysis therefore reinforces the main
picture emerging from the previous results: DLinear provides consistently
low errors across both trajectories and local observables, whereas the
Transformer combines accurate predictions for a substantial fraction of the
test set with much larger errors for a subset of initial conditions.

\section{Discussion and conclusions}

Our results reveal a clear contrast between linear and nonlinear forecasting
architectures for the observable dynamics of the PXP model. Despite its
simple structure, DLinear provides accurate and comparatively uniform
predictions across the entire family of initial states considered. The
Transformer, in contrast, exhibits a pronounced dependence on the
initial-state parameter $\theta$: its performance is comparable to that of
DLinear in the region closer to the homogeneous-state limit, but progressively
deteriorates as the initial states approach the N\'eel limit and the dynamics
become increasingly influenced by quantum many-body scarring. In both cases,
the forecasting performance shows only a weak dependence on $\lambda$.

The different error analyses presented above provide a consistent picture of
this behavior. DLinear maintains a low MAE throughout the initial-state
parameter space, while its comparatively low directional accuracy on the raw
predictions originates primarily from small-amplitude, high-frequency
oscillations. Once these short-timescale components are filtered out, the
directional accuracy increases substantially, indicating that the underlying
temporal trend is nevertheless well reproduced. The Transformer exhibits a
qualitatively different behavior: in the region approaching the N\'eel limit,
its errors become larger and more persistent, and the degradation remains
visible after smoothing. This distinction is also reflected in the
feature-resolved error distributions, where the Transformer develops a tail
of trajectories with substantially larger errors, whereas DLinear remains
more uniform across both initial conditions and local observables.

The robust performance of DLinear is particularly notable in view of the
different dynamical regimes explored within the same Hamiltonian. The family
of initial states considered here continuously connects dynamics closer to
the ergodic regime with non-ergodic dynamics characterized by coherent
revivals associated with quantum many-body scars. Nevertheless, the
observable time series considered in this work can be forecast with high
accuracy using a simple linear mapping from a finite window of past
observations to their subsequent evolution. Importantly, this should not be
interpreted as implying that the underlying quantum dynamics are themselves
linear. Rather, it shows that the complexity of the many-body evolution need
not translate directly into the complexity of the mapping required to
forecast a selected set of observables over the time scales considered here.

Conversely, the greater expressiveness of the Transformer does not provide a
systematic advantage in this forecasting task. In particular, its performance
deteriorates as the dynamics approach the quantum-scarred regime, despite the
presence of coherent and recurrent temporal structures in this region. This
result reinforces the observation, already made in classical time-series
forecasting~\cite{zeng2023transformers}, that more expressive nonlinear
architectures do not necessarily outperform appropriately designed linear
models. Our results extend this comparison to observable time series generated
by quantum many-body dynamics and show that the relative performance of the
two approaches can depend strongly on the dynamical regime being forecast.

The present analysis is restricted to a finite PXP chain, a specific family
of initial states, and a selected set of observables and forecasting
horizons. Establishing how broadly the observed behavior extends beyond this
setting therefore remains an open question. A natural direction for future
work is to consider other many-body Hamiltonians exhibiting qualitatively
different ergodic and non-ergodic regimes, in order to determine whether the
robust performance of linear forecasting persists more generally. Extending
the framework to digital quantum circuits would provide another setting in
which to investigate the relation between the complexity of quantum dynamics
and that required for forecasting their observable signatures. Finally,
comparing Markovian and non-Markovian quantum dynamics could help clarify how
memory effects influence the relative performance of linear and nonlinear
forecasting architectures.

\section*{Acknowledgements}
This work has been funded by MIMI (Ministero delle Imprese e del Made in Italy) with the project STROKE 5.0 (CUP B29J23000430005) funded by MIMI Program num. F/310031/01-05/X56, and by the PNRR MUR project PE0000023-NQSTI via the secondary project ``ThAnQ''.

\newpage


\bibliographystyle{unsrt}
\bibliography{sn-bibliography}

\end{document}